\documentclass[usenatbib,usenatbib]{mnras}
\usepackage{todonotes}
\usepackage{times}
\usepackage{amssymb}
\usepackage{amsmath}
\usepackage{graphicx} 
\usepackage{bmpsize}
\usepackage{epstopdf}
\usepackage{dblfloatfix}  
\usepackage[normalem]{ulem}

\newcommand{\HI}{\rm H~{\sc i }}
\newcommand{\HII}{\rm H~{\sc ii }}

\newcommand{\TB}{\delta T_{\rm b}}

\newcommand{\TS}{T_{\rm S}}
\newcommand{\TK}{T_{\rm K}}
\newcommand{\TCMB}{T_{\gamma}}
\newcommand{\lya}{\rm {Ly{\alpha}}}
\newcommand{\lyb}{\rm {Ly{\beta}}}
\newcommand{\OmegaB}{\Omega_{\rm B}}
\newcommand{\Omegam}{\Omega_{\rm m}}

\title[Impact of $\lya$ heating on the global 21-cm signal]{Impact of $\lya$ heating on the global 21-cm signal from the Cosmic Dawn}

\author[Ghara et al.]{
Raghunath Ghara$^{1,2,3}$\thanks{E-mail: ghara.raghunath@gmail.com}, Garrelt Mellema$^{1}$
\\
$^1$ The Oskar Klein Centre, Department of Astronomy, Stockholm University, AlbaNova, SE-10691 Stockholm, Sweden\\
$^{2}$Department of Natural Sciences, The Open University of Israel, 1 University Road, PO Box 808, Ra'anana 4353701, Israel \\
$^{3}$Department of Physics, Technion, Haifa 32000, Israel
}

\date{Accepted XXX. Received YYY; in original form ZZZ}

\pubyear{2019}

\begin{document}
\label{firstpage}
\pagerange{\pageref{firstpage}--\pageref{lastpage}}
\maketitle

\begin{abstract}
The resonance scattering of $\lya$ photons with neutral hydrogen atoms in the intergalactic medium not only couples the spin temperature to the kinetic temperature but also leads to a heating of the gas. We investigate the impact of this heating on the average brightness temperature of the 21-cm signal from the Cosmic Dawn in the context of the claimed detection by the EDGES low-band experiment. We model the evolution of the global signal taking into account the $\lya$ coupling and heating and a cooling which can be stronger than the Hubble cooling. Using the claimed detection of a strong absorption signal at $z\approx 17$ as a constraint, we find that a strong $\lya$ background is ruled out. Instead the results favour a weak $\lya$ background combined with an excess cooling mechanism which is substantially stronger than previously considered. We also show that the cooling mechanism driven by the interaction between millicharged baryons and dark matter particles no longer provides a viable explanation for the EDGES result when $\lya$ heating is taken into account.  
\end{abstract}




\section{Introduction}
\label{sec:intro}
The formation of the first sources of light is one of the milestone events in the history of our Universe. These primordial sources changed the ionization and thermal state of the gas in the intergalactic medium (IGM) and thus affected the further evolution of the Universe. The period when these very first sources formed is sometimes called the `Cosmic Dawn' (CD). Details regarding these early sources, such as the time of their formation, their emission properties, etc. remain unknown. Models such as in \citet{2006MNRAS.372.1093F, Mesinger2013, 2017MNRAS.464.3498F,  2018MNRAS.478.2193C, 2019MNRAS.484..933P, 2019MNRAS.483.1980M},  suggest that they formed around redshift 30 and their ultraviolet radiation first caused the spin temperature of the neutral hydrogen in the IGM to change due to the repetitive scattering of Lyman series photons, a process known as the Wouthuysen-Field effect \citep{wouth52, field58, hirata2006lya, 2006ApJ...651....1C}. The same models also predict that over time X-rays produced by these sources started to heat the IGM and only much later, in what usually is called the Epoch of Reionization (EoR) sufficient numbers of ionizing photons were produced to reionize the Universe.
  
The 21-cm signal produced by the neutral hydrogen in the IGM during these epochs can provide us with answers to many of the questions regarding the CD and the EoR. Therefore several efforts to detect this signal have been initiated. Two different types of experiments exist. The first type uses large interferometers to measure the spatial fluctuations of the neutral hydrogen (\HI) signal in terms of statistical quantities such as the power spectrum. Examples of these are the Low Frequency Array (LOFAR)\footnote{http://www.lofar.org/} \citep{van13, 2017ApJ...838...65P}, the Giant Metrewave Radio Telescope (GMRT)\footnote{http://www.gmrt.tifr.res.in}\citep{ghosh12, paciga13}, the Precision Array for Probing the Epoch of Reionization (PAPER)\footnote{http://eor.berkeley.edu/} \citep{parsons13} and the Murchison Widefield Array (MWA)\footnote{http://www.mwatelescope.org/} \citep{bowman13, tingay13}. The future low-frequency component of the Square Kilometre Array (SKA-Low)\footnote{http://www.skatelescope.org/} will have the sensitivity to directly probe the spatial structure of the fluctuations by producing images of the signal \citep{2015aska.confE..10M, ghara16}. 

The second type of experiment tries to detect the sky-averaged 21-cm signal, a quantity which the interferometers are unable to measure. Such a measurement only requires a single antenna. Examples of this type are EDGES \citep{2010Natur.468..796B}, SARAS \citep{2015ApJ...801..138P},  BigHorns  \citep{2015PASA...32....4S}, SciHi  \citep{2014ApJ...782L...9V} and LEDA \citep{2012arXiv1201.1700G}.  

The detection of the redshifted 21-cm signal from the EoR and CD is very challenging for all types of experiments as it is several orders of magnitude weaker than the galactic and extra-galactic foreground signals at these frequencies. In addition, long integration times are needed to bring the system noise below the cosmological signal which makes calibration challenging, not only because of instrument stability but also because of the impact of time-dependent ionospheric effects. As a consequence, no undisputed detections of the signal have yet been made.

The strength of the redshifted 21-cm signal from the CD depends on the gas temperature and background $\lya$ flux which are determined by the radiation sources and the heating/cooling processes. Several heating processes such as X-ray heating \citep{Pritchard07, Mesinger2013, ghara15b, ghara15c, 2018arXiv180803287R, 2019MNRAS.487.2785I}, shock heating \citep{2004ApJ...611..642F} and heating due to resonance scattering of $\lya$ photons (hereafter `$\lya$ heating') \citep{2004ApJ...602....1C, 2007ApJ...655..843C, 2006MNRAS.372.1093F} can increase the kinetic temperature of the gas in the IGM during these epochs. However, the relative contribution of these mechanisms is uncertain. 
In addition to these heating processes based on known physics, unknown physics such as dark matter decay may also convey energy to the IGM {\citep{2018PhRvD..98d3006C, 2018JCAP...05..069M, 2018arXiv180309739L}}. The gas cooling is expected to be dominated by the adiabatic cooling due to the expansion of the Universe (`Hubble cooling') with radiative cooling due to e.g.\ recombinations playing a subdominant role.

Recently, \citet{EDGES2018} claimed a detection of a redshift-amplitude profile of the global 21-cm signal around redshift $z\sim 17$ from observations with the EDGES low-band instrument. However, the measured absorption signal was found to be stronger by several factors than the signal predicted by the previous theoretical studies such as \citet{Pritchard07, Mesinger2013, santos08, ghara15a}. Explanations for the EDGES low-band results fall into two categories. The first kind assumes a lower than expected IGM temperature due to excess cooling caused by an unknown physical process such as the interaction between baryons and dark matter particles \citep{2018Natur.555...71B, 2018PhRvL.121a1101F, 2018arXiv180210094M, PhysRevLett.121.011102}. The second type considers the presence of an excess radio background which can also enhance the measurement of the \HI ~signal, which is otherwise seen against the background of the Cosmic Microwave Background (CMB)  \citep{2018ApJ...858L..17F, 2018ApJ...868...63E, 2018PhLB..785..159F}.  Examples of sources which could cause such an excess radio background are supermassive black holes \citep{2018ApJ...868...63E} or supernova from first stars at $z\gtrsim 17$ \citep{2019MNRAS.483.1980M}. However, models relying on such astrophysical sources are unlikely as the time scale for generating a radio background is several orders of magnitude shorter than the duration of the EDGES signal centred at redshift $\sim 17$ \citep{2018MNRAS.481L...6S}. In addition the required excess background requires a $\sim 10^3$ times stronger flux of 1-2 GHz photons than observed from local galaxies \citep{2019MNRAS.483.1980M}. The viable alternative is an excess radio background of cosmological origin, e.g. decay of unstable particles into dark photons with non-zero mixing angle with electromagnetism \citep[see e.g.,][]{2018PhRvL.121c1103P}.

Both of these explanations require the spin temperature to be strongly coupled to the gas temperature which in turn requires a strong $\lya$ background. However, these $\lya$ photons will also heat up the gas by resonance scattering. The question is whether this heating effect has an impact on the global signal. \citet{madau1997} estimated the heating rate due to resonance scattering assuming that the scatterings occur with atoms at rest. For this estimate, the IGM temperature would exceed the CMB temperature in a fraction of Hubble time. A subsequent paper by \citet{2004ApJ...602....1C} included the effect of atomic thermal motions into the calculation and showed that the $\lya$ heating rate is at least three orders of magnitude lower than estimated in \citet{madau1997}. Their calculation considered heating due to photons between $\lya$ and $\lyb$ (so-called `continuum photons') as these redshift into the $\lya$ resonance and cooling due to the cascade of higher resonance states into $\lya$ (so-called `injected photons'). These authors showed that these two mechanism balance at a temperature $\sim 10$ K and thus the temperature would not increase beyond that. This low equilibrium value has prompted many works to neglect $\lya$ heating as its effect would seem to be negligible compared to for example X-ray heating. However, both these works did not consider the forbidden transition from the 2s to the 1s level of hydrogen, something which was added to the calculation by \citet{2007ApJ...655..843C} who furthermore included the effect of deuterium. The result is a lower cooling contribution from the injected photons and which implies that the gas temperature can increase to an equilibrium value of $\sim 100$ K prior to the reionization. 

Previous studies of the global 21-cm signal in the context of the EDGES results did either not consider $\lya$ heating {\citep[see e.g.,][]{PhysRevLett.121.011103, PhysRevD.98.023501, PhysRevD.98.063021, 2018arXiv180303091B, 2018arXiv181209760N}} or used the erroneously low values from \citet{2004ApJ...602....1C}
{\citep[see e.g.,][]{2019arXiv190202438F, 2018PhRvD..98j3513V}}.
In this study, we for the first time adopt the calculation of the $\lya$ heating rates from \citet{2007ApJ...655..843C} and investigate its impact on the global 21-cm signal from the CD. We include excess cooling so as to be able to reproduce EDGES low-band observations. We will consider models which use an excess radio background in a future work. We explore the parameter space of $\lya$ heating and excess cooling to study the absorption profile of the global signal to find combinations of parameters that agree with the EDGES low-band results.

We have organised the paper in the following way. In Section \ref{sec:model} we describe the analytical model we use to calculate the evolution of the global 21-cm signal, including the heating rates due to resonance scattering of the $\lya$ photons. We first present results for a phenomenological excess cooling rate in Section \ref{sec:result}, followed by an investigation of a physically motivated excess cooling rate in Section~\ref{sec:DMB}. We conclude in Section \ref{sec:con}. Throughout the paper we use the following set of cosmological parameters $\Omegam=0.32$, $\OmegaB=0.049$, $\Omega_\Lambda=0.68$, $h=0.67$, $\sigma_8=0.83$ and $n_{\rm s}=0.96$ \citep{Planck2015}.


\section{Model for 21-cm signal}
\label{sec:model}

\subsection{Analytical model}
\label{sec:analytical}
The 21-cm signal from the \HI gas is measured as the differential brightness temperature against the CMB and can be written as
\begin{equation}
 \TB   =  27 ~x_{\rm HI} (1+\delta_{\rm B}) \left(\frac{\OmegaB h^2}{0.023}\right) \sqrt[]{\frac{0.15}{\Omegam h^2}\frac{1+z}{10}}  \left(1-\frac{\TCMB}{\TS} \right)\,\rm{mK},
\label{eq_tb}
\end{equation}
where $x_{\rm HI}$, $\delta_{\rm B}$, $\TS$ and $\TCMB=2.73 \times (1+z)$ K denote the neutral fraction, density contrast, spin temperature of the hydrogen gas and CMB temperature at redshift $z$, respectively. 

We adopt an analytic approach to model the expected 21-cm signal in the presence of spin temperature fluctuations. This approach follows previous works such as \citet{Pritchard07, 2005ApJ...630..643M}. It incorporates $\lya$, UV and X-ray photons from the sources which are taken to be associated with dark matter halos. The number of dark matter halos at a given redshift is determined using the Press-Schechter halo mass function. We assume that only halos with virial temperatures above $10^4$ K contribute. The model estimates the volume averaged ionization fractions of the highly ionized \HII ~regions ($x_i$) and of the mostly neutral gas in the IGM outside these \HII ~regions ($x_e$). We assume the temperature of the ionized \HII ~regions to be $\sim 10^4$~K. The gas temperature ($\TK$) of the largely neutral medium outside the \HII ~regions is calculated using the various heating and cooling processes.

The heating rate due to resonance scattering as well as the spin temperature coupling depend critically on the number of $\lya$ photons emitted from  the sources. To estimate the average $\lya$ photon flux, we follow the method from \citet{2006MNRAS.372.1093F}. We assume a power law spectrum  $\epsilon_s(\nu) = f_{\alpha} A_{\alpha} \nu^{-\alpha_{s}-1}$ between $\lya$ and $\lyb$ and between $\lyb$ and the Lyman limit, where the power law indices $\alpha_{s}$ can differ. The spectral index $\alpha_s$ between $\lya$ and $\lyb$ is taken to be 0.14 which corresponds to population II type sources. The normalization factor $A_{\alpha}$ is estimated such that the number of $\lya$ photons per baryon in the range $\lya$-$\lyb$ is 6520 for $f_{\alpha}=1$. The spectral index in the range $\lya$-Lyman limit is adjusted so that the total number of photons per baryon for this wavelength regime is 9690. The parameter $f_\alpha$ determines the production rate of the $\lya$ photons from the stars. The heating rate due to $\lya$ scattering is described below in Section~\ref{sec:heat}.

To model the X-ray heating, we follow \citet{Pritchard07} and assume that the emissivity of X-ray photons from the sources follows the star formation rate density. We use an X-ray spectral distribution given by \begin{equation}
\epsilon_X(\nu) = \frac{L_0}{h\nu_0}\left(\frac{\nu}{\nu_0}\right)^{-\alpha_X-1}\,,
\end{equation}
with $L_0=f_X\times 10^{41} ~\rm erg ~s^{-1} ~Mpc^{-3}$, $h\nu_0 = 1 ~\rm keV$. For our fiducial X-ray source we choose the X-ray efficiency parameter to be $f_X=1$ and the spectral index of the X-ray spectrum to be $\alpha_X=0.5$. Note that for most of our results we will set $f_X=0$ as we want to focus on the effect of $\lya$ heating. This makes our results conservative as additional X-ray heating will only further increase the gas temperature.

Finally our model also includes the effect of ionizing UV radiation. The rate of emission of the UV photons per baryon is
\begin{equation}
\Lambda_i = \zeta \frac{ {\rm d} f_{\rm coll}}{ {\rm d} t}.
\end{equation}
The ionization efficiency parameter $\zeta = N_{\rm ion}\times f_{\rm esc}\times f_{\star}$ depends on the average number of ionizing photons per baryon produced in the stars ($N_{\rm ion}$), the star formation efficiency ($f_\star$) and the escape fraction of the UV photons ($f_{\rm esc}$). All these quantities are uncertain during the CD and EoR. In this study, we assume $N_{\rm ion}=4000$ which corresponds to population II types of stars, $f_\star=0.1$ and $f_{\rm esc}=0.1$ for modelling reionization. We note however that for most of our results ionization levels remain very low and do not impact the global 21-cm signal.

\begin{figure*}
\begin{center}
\includegraphics[scale=0.8]{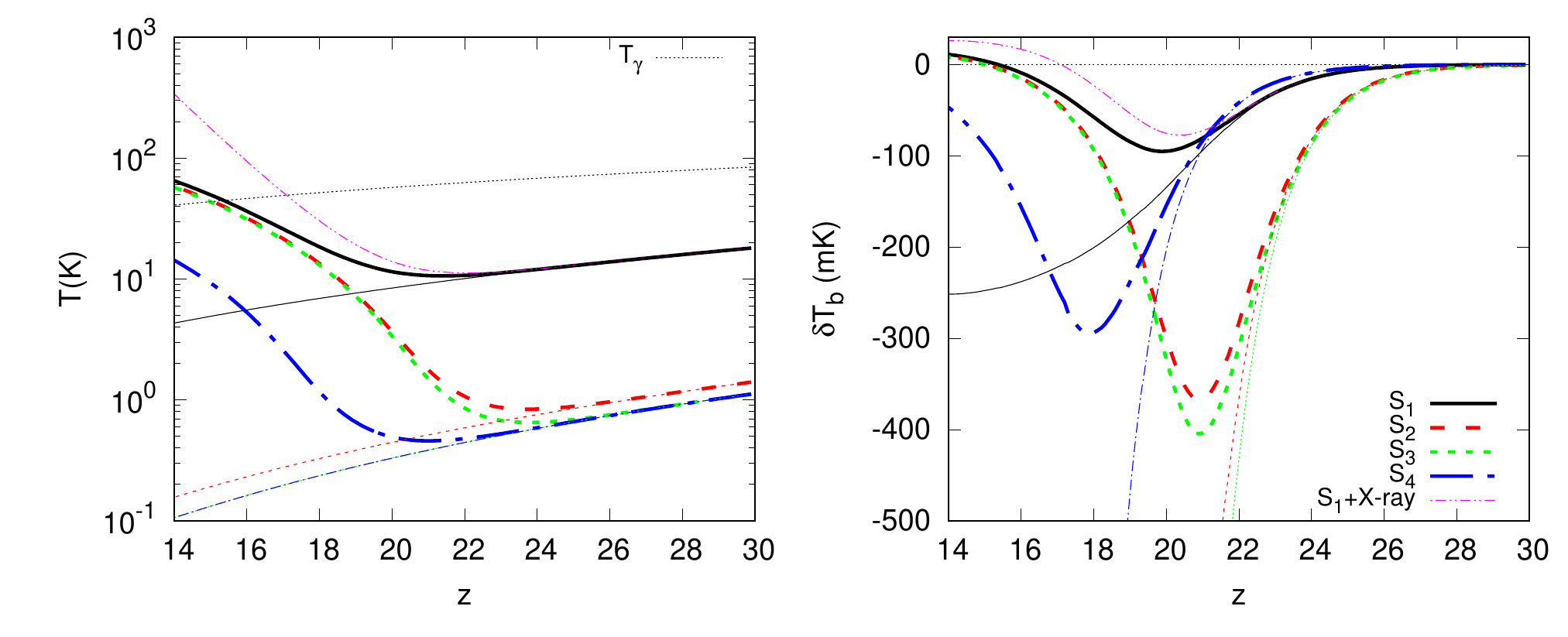}
    \caption{Left-hand panel: The redshift evolution of the gas temperature for the different scenarios described in Section~\ref{sec:main_scene}, also see Table~\ref{tab2}. The thin lines show the cases without $\lya$ heating, the thick lines the cases with $\lya$ heating. The double dot-dashed curve shows the case of model S$_1$ with X-ray heating but without $\lya$ heating. The black dotted curve indicates the evolution of the CMB temperature with redshift. Right-hand  panel: The redshift evolution of the volume averaged differential brightness temperature $\TB$ for the same scenarios.}
   \label{image_global}
\end{center}
\end{figure*}

\subsection{Heating due to resonance scattering}
\label{sec:heat}
To estimate the heating rates due to the resonance scattering, we follow the calculations of \citet{2007ApJ...655..843C}. Photons emitted with frequencies between $\lya$ and $\lyb$ frequency (`continuum photons') will redshift to the $\lya$ frequency at which point they suffer resonance scattering by \HI. This process will heat up the gas. {On the other hand, photons with a wavelength between $\lyb$ and Lyman limit will be absorbed by the hydrogen atoms after redshifting to $\lyb$ or other higher Lyman series lines. If higher resonance or excited states first decay to $2p$ state and then to the ground state, one $\lya$ photon will be emitted.   } In contrast to the continuum photons, the emission of the $\lya$ photons (`injected photons') due to the cascade of from the higher levels will cool the gas. The spectrum gets affected once the photons redshift through the $\lya$ resonance. The intensity $J(\nu)$ at a frequency $\nu$ in the vicinity of the resonance frequency $\nu_\alpha$ can be written as  \citep{2007ApJ...655..843C},
\begin{equation}
J(x) = J(0) e^{-\frac{2\pi \gamma x^3}{3a}-2 \eta x},
\end{equation}
for the injected photons. The above expression also hold for the continuum photons with $x>0$, otherwise
\begin{equation}
J(x) = 2\pi J_0\gamma a^{-1} \int_{-\infty}^{x} e^{\frac{2\pi \gamma (z^3-x^3)}{3a}+2 \eta (z-x)} dz.
\end{equation}
where
\begin{align}
\label{eqn:eqlabel}
\begin{gathered}
x=(\nu/\nu_\alpha - 1)/(2k_{\rm B}\TK/mc^2)^{1/2},
\\
 a=A_{21}(2k_{\rm B}\TK/mc^2)^{-1/2}/4\pi\nu_\alpha,
 \\
 \gamma = \tau_{\rm GP}^{-1}(1+0.4/\TS)^{-1},
 \\
 \eta = [h \nu_\alpha/(2k_{\rm B} \TK m c^2)^{1/2}][(1+0.4/\TS)/(1+0.4/\TK)].
\end{gathered}
\end{align}
Here $k_{\rm B}, m$, $c$ and $A_{21}$ are the Boltzmann constant, mass of hydrogen atom, speed of light and the Einstein spontaneous emission coefficient of $\lya$ transition respectively. The quantities $\tau_{\rm GP}$ and $J_0$ are the Gunn-Peterson optical depth and the UV intensity at a frequency far away from $\nu_\alpha$, respectively.

The quantity $J(0)$ can be expressed as,
\begin{equation}
\frac{J(0)}{J_0} = \frac{\pi\zeta\left(J_{1/3}(\zeta)-J_{-1/3}(\zeta)\right)}{\sqrt{3}}+ _1F_2\left(1; 1/3, 2/3, -\zeta^2/4 \right)
\end{equation}
where $\zeta = \sqrt[]{16\eta^3a/9\pi\gamma}$, $_1F_2$ is hyper-geometric function, $J_{1/3}$ and $J_{-1/3}$ are the Bessel functions of first kind respectively.

The total heating/cooling rate due to the resonance scattering can be written as,
\begin{equation}
{\frac{d\log \TK}{d\log t}} \bigg|_{\rm heating} = \frac{2 t}{3 k_{\rm B} \TK} H_{\alpha},
\label{eq:heating}
\end{equation}
where $t$ represents time, $H_{\alpha}$ is the rate of exchange of total energy by the photons due to resonance scattering.
\begin{equation}
H_{\alpha} = \dot{N}_{\alpha} \left(\Delta E_c + \frac{J_i}{J_c} \Delta E_i \right)
\end{equation}
where $\dot{N_{\alpha}}$ denotes the number of photons per hydrogen atom that pass through resonance scattering per unit time. The ratio of injected and continuum photons $J_i/J_c$ depends on the source's surface temperature. We choose  $J_i/J_c \approx 0.1$ which corresponds to a source with an effective temperature $\lesssim 5\times10^4$ K which corresponds approximately to population II type of sources \citep{2007ApJ...655..843C}. The quantities $\Delta E_c$ and $\Delta E_i$ are the total energy gain by the gas due to a resonance scattering by the continuum and injected photons respectively. This can be written as,
\begin{equation}
\Delta E(x) =\frac{(h\nu)^2}{m c^2} \int \frac{J(x)}{J_0} \phi(x) dx
\end{equation}
where $\phi(x)$ is the normalized scattering cross-section. Note that \citet{2007ApJ...655..843C} also considered the heating contribution from deuterium in their studies. Here we have not included this and thus our calculations somewhat underestimate the actual heating rates.


\subsection{Cooling processes}
\label{sec:ex-cool}
The gas temperature of the IGM is one of the key quantities which determines the strength and nature of the 21-cm signal from the CD. As the heating and cooling processes during those epochs are uncertain, the gas temperature as well as the signal are poorly understood. The analytical method used in this study incorporates the adiabatic cooling due to the expansion of the Universe which dominates over radiative processes such as the collisional-ionization cooling, recombination cooling, collisional excitation cooling, free-free cooling, etc. After Compton scattering with CMB photons ceases to be important, this Hubble cooling causes the average gas temperature to evolve as $\TK \propto (1+z)^2$. For standard physics, the post-recombination gas temperature is easily calculated, as can for example be done with the publicly available code {\sc recfast} \citep{1999ApJ...523L...1S}. The results show that for our cosmological parameters the $\TK \propto (1+z)^2$ relation is valid below $z_0\approx 138$. Expressed in the same form as the $\lya$ heating rate in Equation~\ref{eq:heating}, this adiabatic or Hubble cooling is given as
\begin{equation*}
  \frac{d\log T_K}{d\log t}\bigg|_{\rm H} = -\frac{4}{3}.
\end{equation*}

However, as pointed out by \citet{EDGES2018} this cooling process is unable to explain the strong absorption signal at redshift 17 found in the EDGES low-band results as it requires a lower temperature than can be achieved using standard cosmological models. In order to reproduce the EDGES results we therefore need to include an excess cooling rate in our calculations. Here we make two choices. In Section~\ref{sec:result} we use a simple phenomenological excess cooling model and in Section~\ref{sec:DMB} we use a physically motivated excess cooling rate based on interactions between dark matter particles and baryons.

\section{Phenomenological cooling model}
\label{sec:result}
In this section, we consider a simple phenomenological cooling rate inspired by \citet{2019MNRAS.483.1980M}, given by
\begin{equation}
\frac{d\log T_K}{d\log t}\bigg|_{\rm cool} = \alpha \left[\frac{1+z}{1+z_0}\right]^\beta\,.
\label{eq:ex-cool}
\end{equation}
The parameters $\alpha\le 0$ and $\beta$ determine the strength and redshift dependence of the excess cooling rate respectively. We only apply this excess cooling rate for redshifts $z\le z_0$. However, in principle, $z_0$ could be treated as a free parameter. 

To gain insight into the impact of $\lya$ heating on the volume averaged 21-cm signal from the CD, we first show the results for a number individual scenarios (Section~\ref{sec:main_scene}). After this we will explore the parameter space made up of $\alpha$, $\beta$ and the average $\lya$ flux (Section~\ref{sec:parameter}). Lastly, we will investigate scenarios that can explain the strong absorption signal as reported by the EDGES low-band observation (Section.~\ref{sec:edges_interpretation}).

\begin{table}
\centering
\small
\tabcolsep 6pt
\renewcommand\arraystretch{1.5}
   \begin{tabular}{c c c c c c c}
\hline
\hline
Scenarios & $f_\alpha$ & $\alpha$ & $\beta$  & ${\TB}_{,\rm min}$ & $z({\TB}_{, \rm min})$ & $\Delta z$	 \\

\hline
\hline
S$_1$ & 1.0   & 0.0   &  0.0     & -95.1	  & 19.9	& 4.6		\\
S$_2$ & 1.0 & -1.0	& 0.0	&	-367.1 & 20.9	& 3.7 \\
S$_3$ & 1.0	& -1.0 & -0.1	&	-404.6 & 20.9 & 3.6 \\
S$_4$ & 0.1	& -1.0 & -0.1	&	-294.4 & 17.9 & 4.2 \\
\hline
\hline
\end{tabular}
\caption[]{The $\lya$ efficiency and excess cooling parameters for the four different scenarios considered in Section~\ref{sec:main_scene}. Also shown are the quantities which describe the resulting absorption profile, ${\TB}_{,\rm min}$, $z({\TB}_{,\rm min})$ and $\Delta z$ which represent the minimum brightness temperature, its corresponding redshift and the FWHM of the absorption profiles, respectively.}
\label{tab2}
\end{table}

\begin{table}
\centering
\small
\tabcolsep 6pt
\renewcommand\arraystretch{1.5}
   \begin{tabular}{c c c c c}
\hline
\hline
Parameters & Min range & Max Range 	 \\

\hline
\hline
$f_\alpha$   & 0.01 & 100.0   			\\
$\alpha$ & -1.5 & 0.0		\\
$\beta$ 	& -0.5 & 0.5  \\
\hline
\hline
\end{tabular}
\caption[]{The range of the three parameters for the phenomenological cooling model explored in this study. }
\label{tab1}
\end{table}

\begin{figure*}
\begin{center}
\includegraphics[scale=0.43]{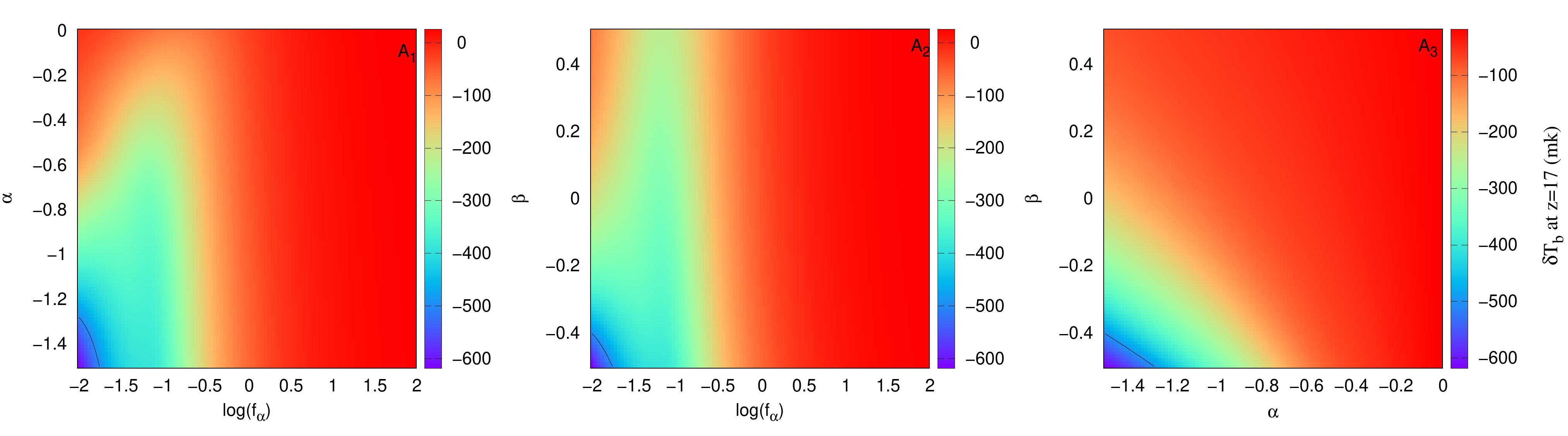}
    \caption{The different panels show the averaged brightness temperature at redshift 17. We vary two parameters at a time in these slices, the third parameter is fixed such that each slice contains the lowest brightness temperature at redshift 17 for the entire parameter space. The third parameter values are $\beta=-0.5$, $\alpha=-1.5$ and $f_\alpha=0.01$ for panels $A_1$, $A_2$ and $A_3$, respectively. The black contours represent -500 mK brightness temperature as reported by the EDGES low-band observation. }
   \label{image_paramz}
\end{center}
\end{figure*}

\begin{figure*}
\begin{center}
\includegraphics[scale=0.46]{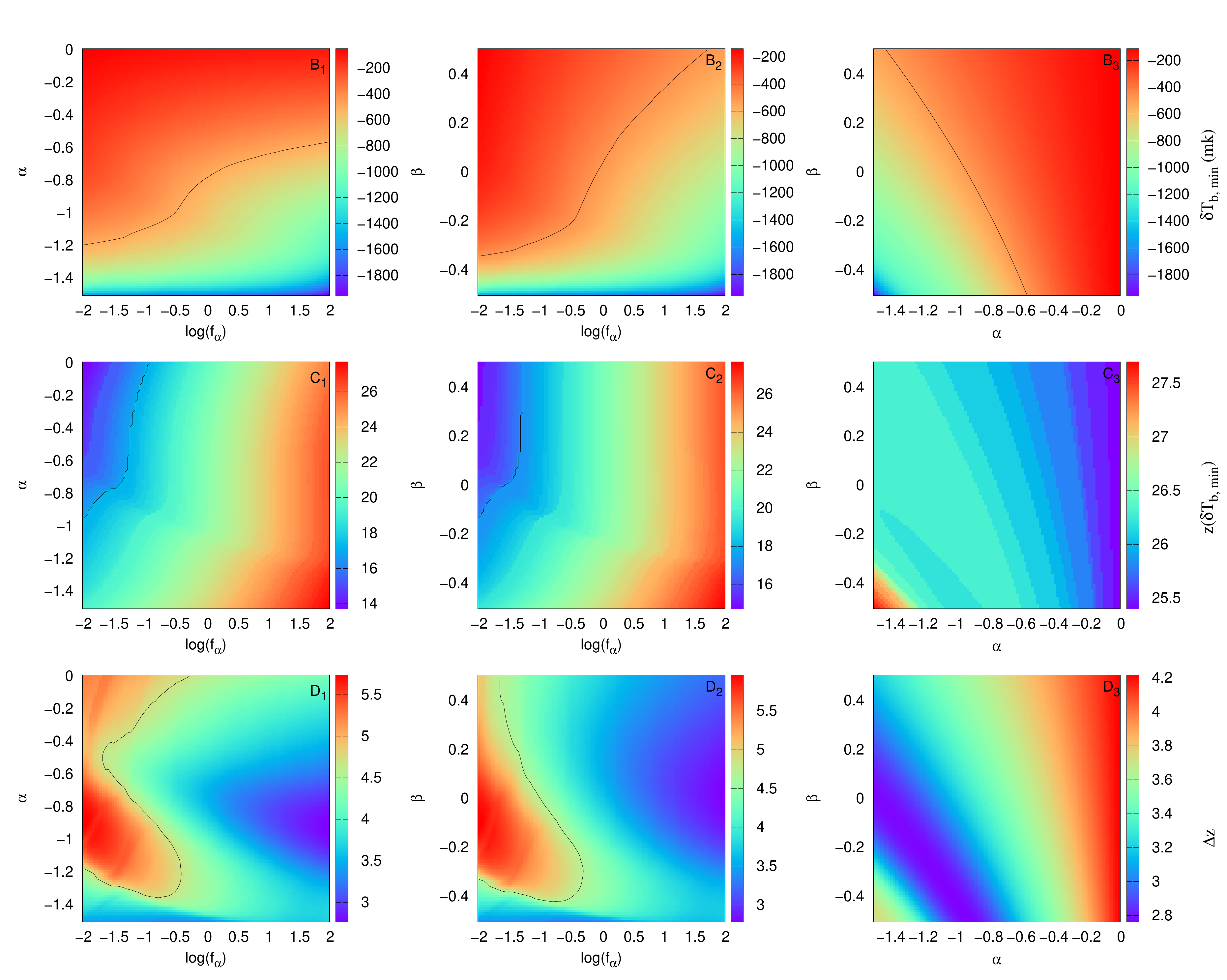}
    \caption{The upper panels represent the minimum brightness temperature throughout Cosmic Dawn ${\TB}_{,\rm min}$ in 2D slices through the parameter space. The third parameter is fixed such that the slices contain the lowest brightness temperature obtained while exploring the entire 3D parameter space. In this case, the third parameter values are $\beta=-0.5$, $\alpha=-1.5$ and $f_\alpha=100$ which correspond to panels $B_1$, $B_2$ and $B_3$ respectively. The middle row panels represent the associated redshifts, $z({\TB}_\mathrm{, min})$ and the bottom panels the FWHM $\Delta z$ of the absorption profile of these models. The contours from top to bottom panels of the figure represent ${\TB}_{,\rm min}=-500$ mK, $z({\TB}_\mathrm{, min})=17.2$ and $\Delta z=4.7$ respectively which characterize the absorption profile as reported by the EDGES low-band observation.}
   \label{image_param}
\end{center}
\end{figure*}

\subsection{Exploratory scenarios}
\label{sec:main_scene}
We choose four different sets of parameters to study the impact of different parameters/processes on the evolution of $\TB$. The parameters for these scenarios are listed in Table \ref{tab2}. The fiducial model S$_1$ has $\alpha=0$ and therefore does not include any excess cooling. The left-hand panel of Fig.~\ref{image_global} presents the redshift evolution of the average gas temperature of the neutral regions in the IGM for these four choices. For each, we consider two cases, namely without (thin lines) and with (thick lines) $\lya$ heating. For the choice of no excess cooling S$_1$ we also consider a case without $\lya$ heating but with heating by X-ray sources (thin double dot-dashed curve).

For the cases without $\lya$ and X-ray heating, the temperature keeps decreasing over time as no other heating mechanisms are included in these scenarios. When we include $\lya$ heating, it impacts the gas temperatures as early as redshift 22 in all these scenarios. For the scenario without excess cooling, the gas temperature increases to $\sim 60$ K at redshift $\sim 14$ which is roughly consistent with the results of \citet{2007ApJ...655..843C}. The small difference is due to ignoring the contribution from deuterium in our calculations. When we instead of $\lya$ heating include X-ray heating according to the description in Section~\ref{sec:analytical}, the gas temperature for S$_1$ increases more rapidly and reaches $\sim 300$~K by $z \sim 14$. This is why $\lya$ heating is often ignored in simulations as X-ray heating will quickly dominate. However, if X-ray heating is inefficient or absent, $\lya$ heating will have a non-negligible impact on the IGM temperature.
 
As scenarios S$_2$--S$_4$ include excess cooling, the Cosmic Dawn starts at lower gas temperatures than for S$_1$. In S$_2$ the excess cooling does not have a redshift dependence, in S$_3$ it increases with time. S$_4$ has the same excess cooling parameters as S$_3$ but a ten times lower $\lya$ efficiency. When including the heating due to the scattering of $\lya$ photons, it impacts the temperatures in S$_2$ and S$_3$ earlier compared to S$_1$, although the background $\lya$ flux densities for these models are identical. This is due to the fact that the $\lya$ heating rates increases as the kinetic temperature decreases (see Equation~\ref{eq:heating}). As expected the heating starts later for a lower $\lya$ background (S$_4$).

Scenarios S$_1$, S$_2$ and S$_3$ each start with different temperatures. However, by $z\sim 16$ they all reach almost the same equilibrium temperature due to $\lya$ heating. For $\alpha=1$ the excess cooling can thus not compete with $\lya$ heating. For the case of a lower $\lya$ flux (S$_4$), the heating is delayed and remains weaker compared to the other scenarios.

The right-hand panel of Fig.~\ref{image_global} shows the redshift evolution of the global 21-cm signal corresponding to the nine scenarios (S$_1$ through S$_4$ with and without $\lya$ heating and S$_1$ with X-ray heating). Note that we always include the $\lya$ coupling for the spin temperature, even in those models where we ignore $\lya$ heating. As for all these scenarios the IGM remains highly neutral at redshifts $>14$, the average brightness temperature is mostly determined by the gas temperature and the strength of the $\lya$ coupling. As the background $\lya$ flux is low at high redshift ($z\sim30$), the coupling between $\TS$ and $\TK$ is weak and $\TS$ remains close to $\TCMB$. This makes $\TB\approx 0$ at those redshifts. As more sources form with time, $\lya$ coupling becomes stronger and the signal starts to appear in absorption, i.e., with a negative sign. However, different heating processes can increase the gas temperature and eventually $\TB$ transitions from absorption to emission. This produces a characteristic trough-like feature in the redshift evolution of $\TB$, which we refer to as the `absorption profile'.

In the absence of $\lya$ or X-ray heating, $\TB$ decreases with redshift as $\TK$ decreases with time and the signal remains in absorption until reionization ends. In such cases, $\TB$ slowly decreases to $\sim$ -250 mK at redshift $\sim 15$ for S$_1$ without excess cooling (thin solid line), while $\TB$ rapidly decreases to values below $\sim -500$ mK at redshift $\lesssim 20$ for models S$_2$ through S$_4$ which include excess cooling (thin long-dashed, short-dashed and dot-dashed lines). 

In the presence of $\lya$ heating, the increase of gas temperature as early as redshift $\sim 20$ resists the decrease of $\TB$ with time and produces prominent absorption profiles (thick lines). The minimum $\TB$ values of these profiles are much less deep than the corresponding signals from the no heating cases, demonstrating the large impact $\lya$ heating has. For example in scenario S$_1$ the absorption profile does not reach below -100~mK and for S$_2$ and S$_3$ not below -400~mK.

The absorption profiles can be described by the minimum value of the brightness temperature (${\TB}_{,\rm min}$), the corresponding redshift $z({\TB}_\mathrm{, min})$ and the full width at half maximum (FWHM) of the absorption profile ($\Delta z$). We list the values for the cases with $\lya$ heating in Table \ref{tab2}. These numbers clearly depend on the excess cooling rate and $\lya$ heating rates. We see the absorption profiles are much stronger and appear earlier for models S$_2$ and S$_3$ than for model S$_1$. This is due to the excess cooling in the former models which results in a lower initial gas temperature compared to S$_1$. The values for the $\Delta z$ are lower when excess cooling is present. 

When the $\lya$ background is lower ($f_\alpha=0.1$, scenario S$_4$) the absorption profile becomes less deep, widens and appears later compared to the scenario which has $f_\alpha=1$ (S$_3$), even though the gas temperature is actually lower. However, the profile can still reach a minimum of $\sim -300$~mK, below what can be achieved without excess cooling. 

For completeness, the right-hand panel of Fig.~\ref{image_global} also shows the differential brightness temperature evolution for the scenario without excess cooling and $\lya$ heating but with X-ray heating (thin dot-dot-dashed curve). Due to the higher temperatures, this absorption profile is less deep and somewhat narrower than the corresponding case with $\lya$ heating (thick solid curve). 

\subsection{Parameter space study}
\label{sec:parameter}
Now we will explore the parameter space of excess cooling ($\alpha$ and $\beta$) and $\lya$ flux ($f_\alpha$) to find the impact on the absorption signal from the CD in terms of absorption profile parameters ${\TB}_{, \rm min}$, $z({\TB}_\mathrm{,\rm min})$ and $\Delta z$. The details of the parameter space are given in Table \ref{tab1}. As the excess cooling is due to unknown processes, the parameter ranges for $\alpha$ and $\beta$ chosen here are somewhat arbitrary. However, as we will see this range covers the most interesting results in terms of the absorption feature and the EDGES low-band results. 

We will study the global 21-cm signal around redshift $\sim 17$ which corresponds to $z({\TB}_\mathrm{, min})$ of the EDGES low-band detection. The different panels of Fig.~\ref{image_paramz} represent the value of the differential brightness temperature at redshift 17 in 2D slices through the 3D parameter space. For these slices, the third parameter is chosen such that these slices contain the lowest brightness temperature at $z=17$ within the explored parameter space. The values are $\beta=-0.5$ (panel A$_1$), $\alpha=-1.5$ (panel A$_2$) and $f_\alpha=0.01$ (panel A$_3$). 

The resonance photons impact the signal in two ways: ($i$) heating due to resonance scattering decreases for a lower background $\lya$ flux, ($ii$) coupling of $\TS$ with $\TK$ decreases for a lower $\lya$ background.  These two effects create the vertical feature in $\TB$ around $f_\alpha \sim 0.1$ in panels A$_1$ and A$_2$. In the presence of significant $\lya$ heating (e.g.\ for $f_\alpha >$1), the amplitude of $\TB$ at redshift 17 remains small for all values of $\alpha$ and $\beta$. As shown in panel A$_3$, strong excess cooling  ($\alpha \sim -1.5$ and $\beta \sim -0.5$) can produce a deep absorption feature but only for a very weak $\lya$ flux, reaching  values as low as $-600$ mK for $f_\alpha \sim 0.01$.    

We note that the color bar associated with the panels of Fig.~\ref{image_paramz} represents $\TB$ at redshift 17, not ${\TB}_\mathrm{, min}$ for the choice of parameters. This figure shows that $\TB \sim -500$ mK at redshift 17 is only possible for a weak $\lya$ background and strong excess cooling rates as shown by the contours in the panels. However, this does not mean that the values of $\TB$ in this figure are equal to ${\TB}_\mathrm{, \rm min}$, the minimum of the absorption profiles. Thus, we can not directly compare these with the EDGES low-band observations. However, we can see that a large part of the parameter space corresponds to $\TB$ values larger than -500 mK and thus, should be excluded by the EDGES observation. We will present a detailed investigation of this in Section~\ref{sec:edges_interpretation}.

First we will investigate the behaviour of absorption profiles over the parameter space. The top row of panels of Fig.~\ref{image_param} show 2D slices of ${\TB}_{,\rm min}$ through the entire parameter space. As in Fig.~\ref{image_paramz}, the third parameter is chosen such that these slices contain the lowest value of  ${\TB}_{,\rm  min}$ obtained within the entire parameter space. In this case, the values for the third parameter are $\beta=-0.5$, $\alpha=-1.5$ and $f_\alpha=100$ which correspond to the left, middle and right panels, respectively. The middle row of panels shows the associated redshift $z({\TB}_{,\rm min})$ of the minimum of the absorption profiles and the bottom row the corresponding FWHM $\Delta z$. 

Panels B$_1$ and B$_2$ show that ${\TB}_{,\rm min}$ decreases with increasing $f_\alpha$ as the coupling between $\TS$ and $\TK$ becomes stronger. However, this also implies that $\lya$ heating becomes efficient earlier and thus the minima of the absorption profiles appear at higher redshifts when increasing $f_\alpha$ (see panels $C_1, C_2$). As shown in panel $B_3$,  ${\TB}_{,\rm min}$ decreases for lower values of $\alpha$ and $\beta$ which corresponds to stronger excess cooling and also in this case  $z({\TB}_\mathrm{,\rm min})$ shifts towards higher redshifts (panel $C_3$). The CD starts with a lower gas temperature for smaller values of $\alpha$ and $\beta$. As the $\lya$ heating rate increases for lower temperatures, $\lya$ heating become efficient earlier for a stronger excess cooling model. These results are consistent with our findings in Section~\ref{sec:main_scene}.

\begin{figure*}
\begin{center}
\includegraphics[scale=0.56]{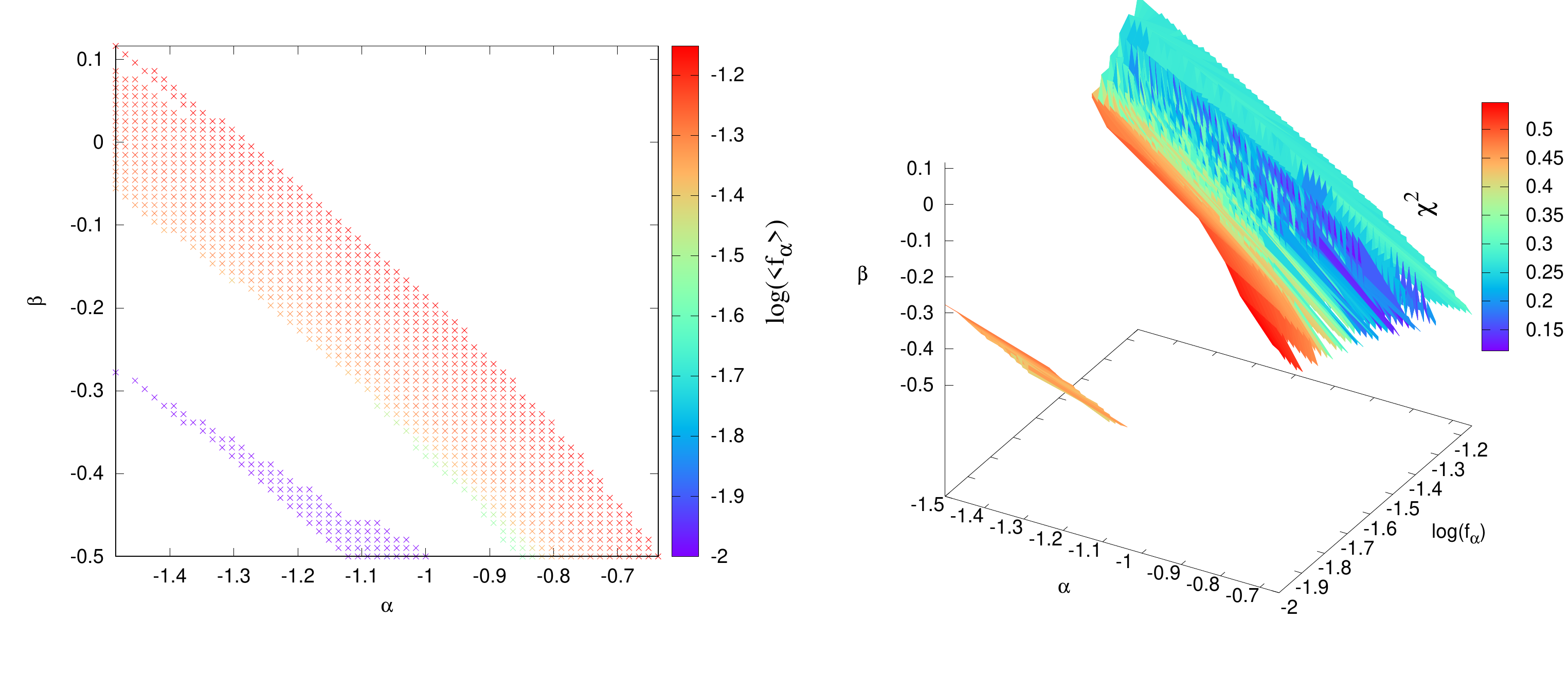}
    \caption{Left-panel: The crosses indicate the values of the excess cooling parameters $\alpha$ and $\beta$ which produce absorption profiles that agree with the EDGES low band results within a $1\sigma$ error. The colour bar represents the average values of $f_\alpha$ for these models.   Right panel: 3D plot of the parameter space which agrees with the EDGES low band results. The colour bar shows the $\chi^2$ error as defined in Equation~\ref{eq:chi2}. }
   \label{image_edges}
\end{center}
\end{figure*}

The bottom row of Fig.~\ref{image_param} shows the corresponding FWHM $\Delta z$. The dependence of $\Delta z$ on the parameters is more complex compared to what we saw for ${\TB}_{,\rm min}$ and $z({\TB}_{,\rm min})$. Here we have to keep in mind a few aspects. One is that all models reach their equilibrium temperature due to $\lya$ heating approximately at the same redshift for a fixed $\lya$ background (as we have seen in Section~\ref{sec:main_scene}). Secondly, the initial temperature (at $z=30$) of these models decreases rapidly with stronger cooling parameters. The absorption profile becomes deeper and shifts towards higher redshift for a larger excess cooling rate. On the other hand, the $\lya$ heating starts earlier and $\TB$ of these profiles approaches zero at a similar redshift. These two facts make $\Delta z$ decrease initially with the increase of excess cooling rate for a fixed $f_\alpha$ as shown in panel $D_3$. However, $\Delta z$ starts increasing for $\alpha < -1$ and $\beta < 0$ as the initial temperature of these models becomes smaller and the minimum of the absorption profiles shifts towards higher redshifts. On the other hand, $\lya$ heating becomes efficient earlier for a larger value of $f_\alpha$ which decrease $\Delta z$ for a fixed excess cooling (see panels $D_1$ and $D_2$).

The black lines in Fig.~\ref{image_param} correspond to the absorption profile parameters estimated from the EDGES low-band observation. However, as these slices correspond to the minimum $\TB$ calculated by exploring the whole 3D parameter space, an interpretation of EDGES results from these contours is difficult. We therefore now turn our attention to the parts of the parameter space that are consistent with the EDGES absorption profile.

\subsection{Interpretation of EDGES low-band results}
\label{sec:edges_interpretation}
\citet{EDGES2018} reported a measurement of a 21-cm absorption profile with ${\TB}_{, \rm min}=-500^{+200}_{-500}$ mK with $z({\TB}_\mathrm{, min})$ and $\Delta z$ equivalent to $78\pm 1$ MHz and $19^{+4}_{-2}$ MHz, respectively. We will investigate what part of our parameter space agrees with this observation. However, we do not consider the detailed shape of the absorption profile as reported in \citet{EDGES2018}, nor use parameter estimation techniques such as Markov chain Monte Carlo. Instead, we consider the values of ${\TB}_{, \rm min}, z({\TB}_\mathrm{, min})$ and $\Delta z$ corresponding to the profile to compare with the absorption profiles produced by our model. We would like to remind the reader that we have not included any X-ray heating in this parameter space study.

Fig.~\ref{image_edges} presents which values for our parameters agree with the EDGES low band results. The left-hand panel shows a 2D plot for parameters $\alpha$ and $\beta$ where the colour of each point represents the average value of $f_\alpha$ for which the values of $\alpha, \beta$  are consistent with the EDGES observation. We see that two specific ranges of cooling parameters produce the desired profile, the broader of the two bands for a weak $\lya$ background flux ($f_\alpha < 0.08$) and the narrower one for a very weak $\lya$ background ($f_\alpha < 0.01$). The broader band is characterized by a strong but not too strong cooling around the redshift of the absorption profile ($-2\lesssim(\mathrm{d}\log T_K/\mathrm{d}\log t)_{\rm cool}\lesssim-1.2)$ and the narrower band by a stronger value of $(\mathrm{d}\log T_K/\mathrm{d}\log t)_{\rm cool}\sim -2.5$ This can also be characterized through the temperature which the IGM would achieve in the absence of $\lya$ heating. For the cooling parameters in the broader of the two bands, this temperature is between 0.2 and 0.3 K and between 0.05 and 0.06 K for the narrower band. For stronger cooling than shown in the left-hand panel, ${\TB}_{, \rm min}$ will be lower and will shift towards higher redshifts. Similarly, $z({\TB}_\mathrm{, min})$ will shift towards higher redshifts for larger values of $f_\alpha$. 

The right-hand panel of Fig.~\ref{image_edges} shows a 3D representation of our parameter space where the colour indicates the $\chi^2$ value. We define $\chi^2$ error in this plot as
\begin{equation}
\chi^2 =\sum_{i=1,3} \left(\frac{M_i-O_i}{\sigma_i}\right)^2,
\label{eq:chi2}
\end{equation}
where $i$ represents there parameters to define the absorption profile used in this study, $M$ and $O$ are the model and observation parameters respectively and $\sigma_i$ represents the $1\sigma$ error on the measured parameters considered here. 
One can notice that for a certain choice of cooling parameters a range of $f_\alpha$ values can satisfy the agreement condition. However, all $f_\alpha$ values are low. The isolated region $f_\alpha=0.01$ corresponds to deeper absorption profiles with ${\TB}_{,\rm min} \lesssim -500 $ mK while the other region 
has absorption depths ${\TB}_{,\rm min} \sim -300 $ mK. 

One thing to keep in mind that we have ignored all other heating processes such as X-ray heating, etc. If any other additional energy is added to the IGM, the excess cooling would have to compensate for this in order for the absorption profile to remain consistent with the EDGES result. In other words, the cooling rates derived here should be considered as lower limits. For example, the combination $f_\alpha \lesssim 0.1$, $\alpha \sim -1.5$ and $\beta \sim 0.1$ corresponds to the minimum excess cooling required to achieve the strong signal reported by EDGES. This minimum excess cooling rate is similar to the Hubble cooling rate at redshift 17.


\section{Physically motivated cooling model}
\label{sec:DMB}
So far we have considered a simple phenomenological form for the cooling rate as given by Equation~\ref{eq:ex-cool}. Now we will consider a physically motivated cooling model based on the interaction between cold dark matter and baryonic particles. Such interactions have the potential to cool the baryonic gas efficiently and explain the EDGES results \citep{2018Natur.555...71B, 2018arXiv180210094M, 2018arXiv180303091B, 2018PhRvL.121a1101F}. However, most of these interaction scenarios are highly constrained by limits from stellar cooling and fifth force experiments. This rules out scenarios in which the cooling of the gas occurs through Rutherford-like scattering with a dominant component of the dark matter. However, a scenario in which cooling is caused by interactions of electrons and protons with a small ($\sim 1\%$) fraction of millicharged dark matter particles is currently not entirely ruled out \citep[see e.g.,][]{2018arXiv180210094M} although only in a very small part of parameter space \citep{2018arXiv180303091B}.

\begin{figure*}
\begin{center}
\includegraphics[scale=0.8]{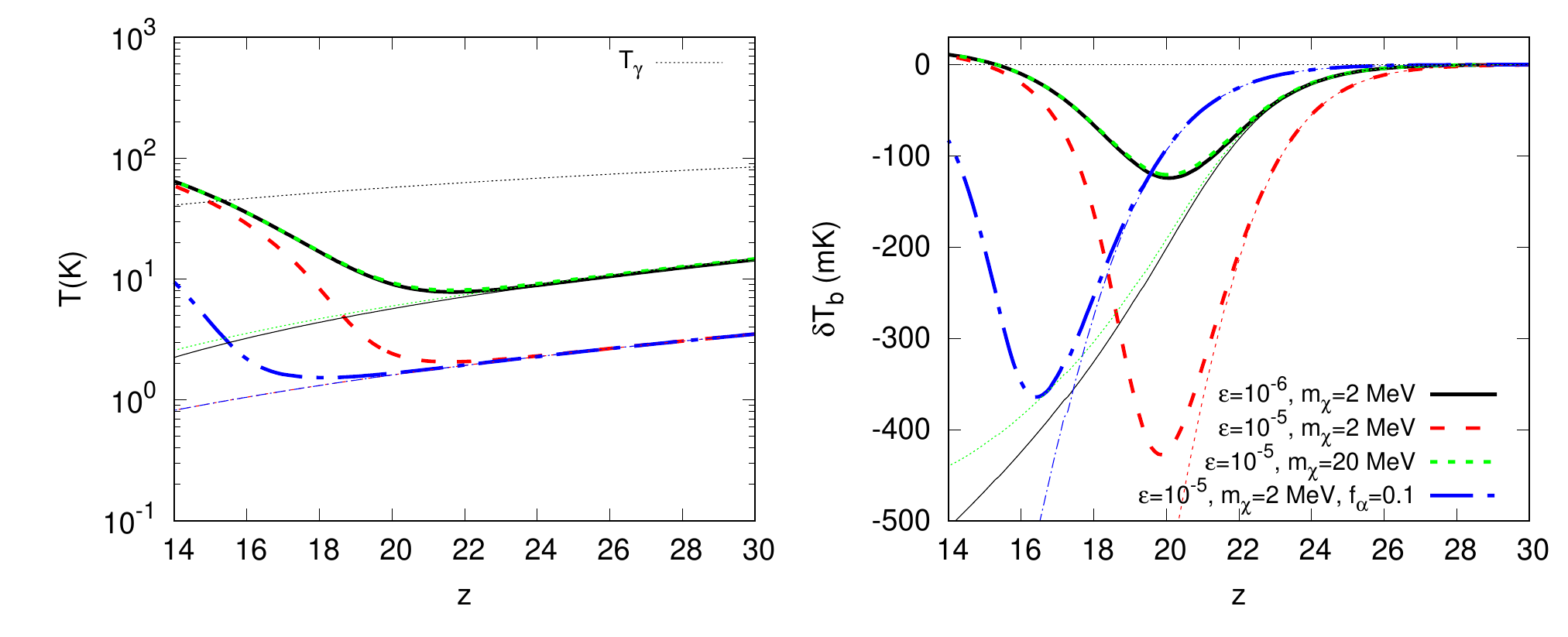}
    \caption{{ Left-hand panel: The redshift evolution of the gas temperature for the different models where cooling of the gas in dominated by interaction with millicharged dark matter. The thin lines show the cases without $\lya$ heating and the thick lines the cases with $\lya$ heating. Different curves correspond to different combinations of $m_\chi$ and $\epsilon$. All curves have been calculated for $f_\alpha=1$ except the thick and thin dot-dashed blue curves which use $f_\alpha=0.1$. The fraction of millicharged dark matter is set to $f_{\rm dm}=0.01$. The black dotted curve in the left panel indicates the evolution of the CMB temperature with redshift. Right-hand  panel: The redshift evolution of the volume averaged differential brightness temperature $\TB$ for the same scenarios.}}
   \label{image_global_DMB}
\end{center}
\end{figure*}

\begin{figure}
\begin{center}
\includegraphics[scale=0.56]{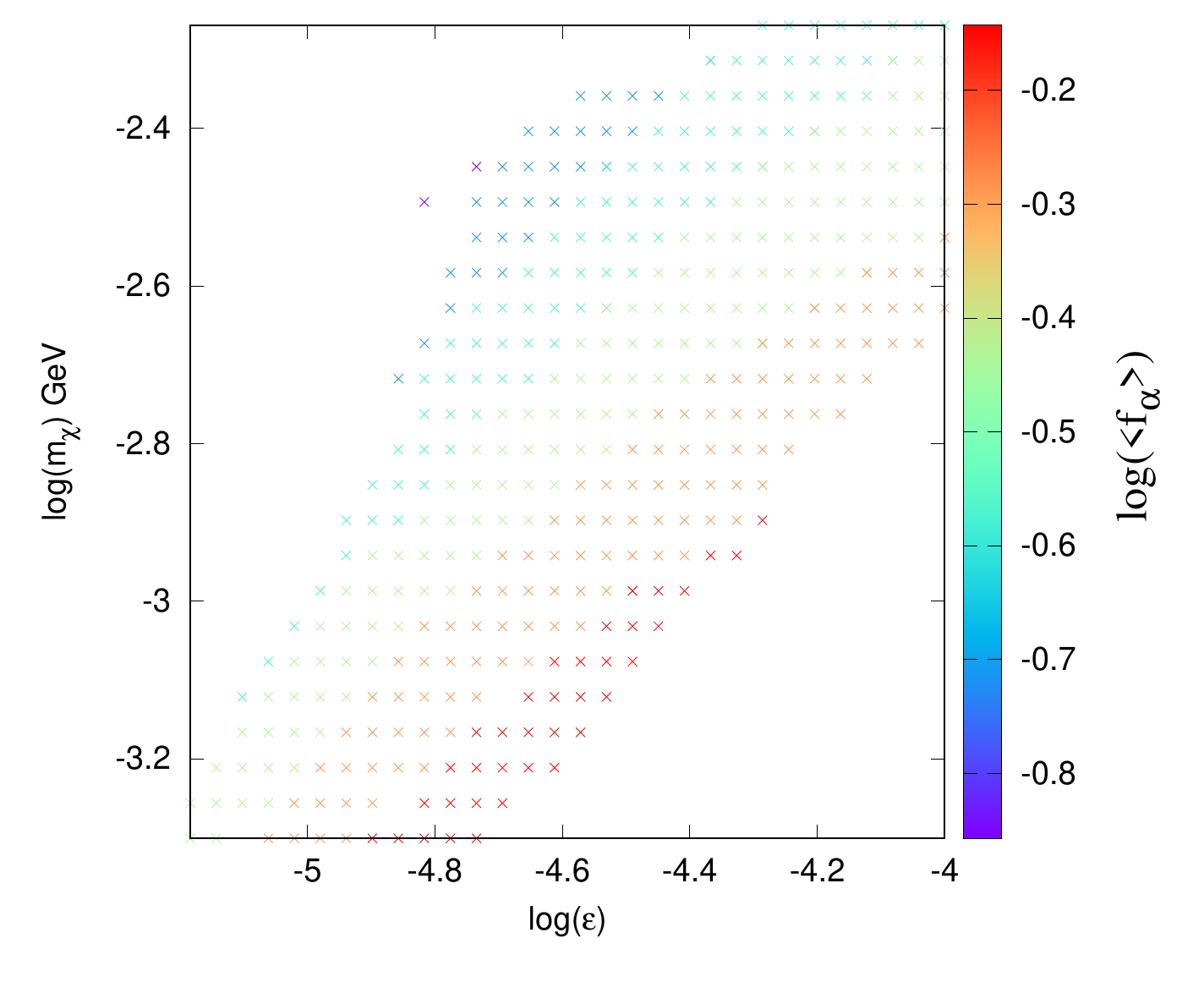}
    \caption{{Parameter study for the millicharged dark matter cooling model. The parameters $\epsilon$ and $m_\chi$ represent the charge and the mass of the dark matter particles. The crosses indicate the values of the parameters which produce absorption profiles that agree with the EDGES low band results within a $1\sigma$ error. The colour bar represents the average values of $f_\alpha$ for these models.} }
   \label{image_edges_DMB}
\end{center}
\end{figure}

The cooling rate for this scenario can be written as
\begin{equation}
\frac{d\log \TK}{d\log t}\bigg|_{\rm cool,DMB} = \frac{4 \dot{Q_b}}{9~H~\TK},
\label{eq:ex-coolDMB}
\end{equation}
where the cooling rate of the baryon $\dot{Q_b}$ can be expressed as the sum of the contributions due to collisions with electrons or protons as targets $t$ \citep{2018arXiv180210094M},
\begin{eqnarray}
\dot{Q_b} \!\!\!\! & = & \!\!\!\! \frac{n_{\chi} x_e}{1+f_{\rm He}}\sum_{t=e,p}\frac{m_t m_{\chi}}{\left(m_t+m_\chi\right)^2}\frac{ \bar{\sigma}_t}{u_{{\rm th},t}} 
\nonumber\\
&\times& \!\!\!\! \left[ \sqrt{\frac{2}{\upi}}\frac{ e^{-r_t^2/2} }{u^2_{{\rm th},t}}\left(T_\chi - \TK\right) + m_{\chi}\frac{F(r_t)}{r_t} \right].
\label{eq:ex-coolDMB1}
\end{eqnarray}
Here, $x_e$ is the residual electron fraction after recombination and $f_{\rm He}\approx0.08$ is the primordial helium fraction. The symbol $m$ stands for mass, where $e$, $p$ and $\chi$ stand for electron, proton and dark matter, respectively. The number density of millicharged dark matter is $n_{\chi} = f_{\rm dm}\times \rho_d/m_{\chi}$ where $\rho_d$ is the dark matter mass density and $f_{\rm dm}$ is the fraction of millicharged dark matter.  $\TK$ and $T_\chi$ represent the temperatures of the baryon gas and the dark matter respectively. The function $F(r_t)$ is defined as 
\begin{equation}
  F(r_t) = {\rm Erf}(\frac{r_t}{\sqrt{2}})-\sqrt{\frac{2}{\upi}}r_t e^{-r^2_t/2}\,,
\end{equation}
where $r_t = v_{\chi b}/u_{{\rm th},t}$, with $v_{\chi b}$ the relative velocity between the baryons and the dark matter and $u^2_{{\rm th},t} = \TK/m_t + T_\chi/m_\chi$ the (iso)thermal sound speed of the DM-t fluid. Finally, the rate also depends on the interaction cross-section between the millicharged dark matter particles and a target $t$, $\bar{\sigma_t}$. We define a charge parameter $\epsilon=e_{\chi}/e$ where $e_{\chi}$ and $e$ are the dark matter and electron charge respectively. The interaction cross-section is assumed to scale with relative velocity 
as $\bar{\sigma_t} = \frac{2\upi \alpha^2_{f}\epsilon^2\xi}{\mu^2_{\chi t}v_{\chi b}^4}$ where $\alpha_{f}$ is the fine-structure constant, $\xi$ is the Debye logarithm and $\mu^2_{\chi t}$ is the reduced mass of the dark matter and target. We refer to \citet{2018arXiv180210094M} for more details on the various terms in Equation~\ref{eq:ex-coolDMB1}.

We follow the approach as in \citet{2018arXiv180210094M} and solve the set of differential equations to track the evolution of the temperatures of the gas and dark matter. We initialize $\TK = \TCMB$ and $T_\chi = 0$ at $z=1010$. We assume that the initial distribution of $v_{\chi b, 0}$ is Gaussian with an root-mean-square value of 29 km~s$^{-1}$. We solve the set of equations for many values $v_{\chi b, 0}$ taken from this distribution and in the end estimate the velocity averaged gas temperatures and brightness temperature.

We will vary two parameters for the cooling rate as described in Equation~\ref{eq:ex-coolDMB}, namely $\epsilon$, the charge of the dark matter particles and their mass $m_{\chi}$. We keep the fraction of millicharged dark matter fixed at $f_{\rm dm}=0.01$ throughout this study.  Figure~\ref{image_global_DMB} shows the redshift evolution of the gas temperature (left panel) and the brightness temperature (right panel) for three different combinations of $m_\chi$ and $\epsilon$. The thin lines show the results with $\lya$ coupling but {\it without} $\lya$ heating and the thick lines include $\lya$ heating. We use $f_\alpha=1$, except for the dot-dashed line which has $f_\alpha=0.1$. These results illustrate the trends associated with the different parameters.

It should first of all be noted that these results are very similar to those shown in Fig.~\ref{image_global}. In absence of $\lya$ heating, a cooling rate with the combination of $m_\chi \sim 2$ MeV and $\epsilon\sim 10^{-6}$ cools the gas to a temperature $\sim 4$ K at redshift 17 which is sufficient to produce a signal which agrees with the EDGES low-band observations, consistent with the results of \citet{2018arXiv180303091B, 2018arXiv180210094M}. However, as expected,  the $\lya$ heating prevents the gas temperature to reach such a low value, even for $f_\alpha=1$. The cooling, as well as the signal, becomes stronger for larger values of $\epsilon$ as this raises the interaction cross-section. On the other hand, increasing $m_{\chi}$ lowers the cooling rate and the signal. As above, we find that the absorption profile shifts towards lower redshifts for lower values of $f_\alpha$. These trends suggest that this cooling model might satisfy the EDGES results for the following two cases: (i) a higher cooling rate than estimated by \citet{2018arXiv180303091B, 2018arXiv180210094M} which can arise due to either a larger $\epsilon$ or a smaller $m_{\chi}$, (ii) a lower $\lya$ flux to keep the heating low at $z\sim 17$.

Next, we vary the parameter $\epsilon$ from $10^{-7}$ to $10^{-4}$ and $m_\chi$ from 0.5 MeV to 1 GeV while we keep the range of $f_\alpha$ the same as used previously. Equivalent to Fig.~\ref{image_edges} for the phenomenological model, Fig.~\ref{image_edges_DMB} presents the parts of parameter space which agree with the EDGES measurements. Note that similar to the earlier case, we have not considered any heating mechanism other than heating due to scattering of the $\lya$ photons. As expected, we find the required $\lya$ flux in this scenario has to be less than 1 as shown by the color bar. Note however that the $f_\alpha$ values found are higher than what we obtained for the phenomenological cooling model. This suggests that the millicharged dark matter cooling process produces a larger cooling rate at redshift $\sim 17$ than the explored range in the previous cooling model.

While \citet{2018arXiv180303091B, 2018arXiv180210094M} conclude that $\epsilon \gtrsim 10^{-6}$ will be required for $m_\chi=2$ MeV to reach agreement with the EDGES results, Fig.~\ref{image_edges_DMB} suggests a larger value of $\epsilon \gtrsim 1.5\times 10^{-5}$ for the same dark matter mass. The required $\lya$ flux for these $m_\chi$ and $\epsilon$ values corresponds to $f_\alpha \sim 0.3$. Clearly, a larger dark matter-baryon interaction cross-section is required when $\lya$ heating is taken into account. 

However, the possible parameter space of the millicharge model in Fig.~\ref{image_edges_DMB} that can explain the EDGES result is disfavoured by the constraints from stellar and super-
nova  cooling,  big  bang  nucleosynthesis  and  a  range  of  particle
physics experiments. Specifically, as shown in fig.~4 in \citet{2018arXiv180303091B}, these constraints require $m_\chi\gtrsim 10$~MeV and for these values we do not find any solutions that are consistent with the EDGES results. We therefore conclude that the millicharged dark matter model no longer offers a viable explanation for the absorption signal claimed by the EDGES team.

\section{Discussions \& Conclusions}
\label{sec:con}
 In this study we have considered the impact of the heating from resonance scattering of $\lya$ photons in the IGM during the Cosmic Dawn on models with excess cooling constructed to explain the deep absorption feature around $z\sim 17$ reported by the EDGES team. This heating is an inevitable result of the resonance scattering which is needed to couple the spin temperature to the gas temperature, the only known process which can produce an observable 21-cm signal from the IGM at these redshifts. The required excess cooling requires new physics and thus its cause remains uncertain. We explored two possibilities, one simple phenomenological form of cooling and one physically motivated one relying on the interaction of putative millicharged dark matter particles with protons and electrons.

 For these two scenarios we investigate the evolution of the average differential brightness temperature of the 21-cm signal. We explore a three-dimensional parameter space defined by two parameters describing the excess cooling ($\alpha$ and $\beta$ for the phenomenological model; $\epsilon$ and $m_\chi$ for the millicharged dark matter) and one parameter setting the strength of the $\lya$ background ($f_\alpha$) to study the global 21-cm signal from the CD. The main findings of the paper are listed below.

Without any excess cooling, $\lya$ heating can start heating the IGM as early as redshift 22 for a typical emissivity of $\sim 10000$ photons per baryon between $\lya$ and the Lyman limit. Although this heating rate is smaller than the usually assumed X-ray heating rates, it can still increase the gas temperature to several tens of K which is the equilibrium temperature between the heating by the continuum photons and cooling by the injected photons. This is consistent with previous studies such as \citet{2007ApJ...655..843C}. For this case, we find an absorption signal of depth $\sim -100$ mK at redshift $\sim 20$.

When including excess cooling, the Cosmic Dawn starts with a very cold IGM. In such cases, $\lya$ heating becomes efficient earlier and rapidly increases the IGM temperature to the equilibrium temperature. For these cases, we find absorption signals which can be factors 3 -- 4 deeper than without excess cooling. 

The exploration of the parameter space of the excess cooling rate parameters and the $\lya$ background shows that the EDGES low-band results can only be reproduced for strong excess cooling combined with a weak $\lya$ background. This puts an upper bound on the background $\lya$ flux which is $\sim$15 times lower than our fiducial choice for the phenomenological cooling model, while for the millicharged dark matter model this upper limit is only a factor $\sim 2$ below the fiducial value. Thus the sources at redshift $\sim 17$ emit fewer $\lya$ photons or the star formation efficiency is lower than expected.  This result disagrees with the findings of \citet{2019MNRAS.483.1980M} who claim that the star formation efficiency should be higher than expected in order to produce the strong $\lya$ background needed to achieve strong coupling between the spin and gas temperatures. However, these authors did not consider the effect of $\lya$ heating.

 Although we find that the millicharged dark matter model can reproduce the EDGES results for a relatively low $\lya$ background and some combinations of DM charge and mass, these combinations are actually ruled out by constraints from stellar and supernova cooling, big bang nucleosynthesis and a range of particle physics experiments \citep{2018arXiv180303091B}. Including $\lya$ heating therefore removes this model as a feasible explanation for the EDGES results.

In our exploration of the parameter space for the phenomenological model we frequently found interesting models at the edge of the parameter ranges that we considered. We did not explore a larger range of values as the trend is quite clear: only fairly strong cooling which without $\lya$ heating would give gas temperatures below $\sim 0.3$~K around $z\approx 17$ combined with a weak $\lya$ background ($f_\alpha \lesssim 0.06$)  
can reproduce the EDGES low-band results. Possibly even stronger cooling with an even weaker $\lya$ background would also give consistent results but such models become increasingly unlikely.

 In general, $\lya$ heating works against all kind of excess cooling models that might explain the EDGES result and will potentially provide strong bounds on their parameters. The same is true for the alternative solutions which rely on a stronger radiation background at the Rayleigh-Jeans tail of the CMB \citep[see e.g.,][]{2018PhRvL.121c1103P}. However, we leave the study of the impact of $\lya$ heating on those types of models to a future study.

We did not explore the impact of changing the source population. In our models, all halos with a virial temperature above $10^4$~K contribute to the $\lya$ background. Obviously increasing this limit would also reduce the background and possibly lead to models in which fiducial values for $f_\alpha$ combined with strong excess cooling could reproduce the EDGES low-band absorption profile. Lowering the minimum virial mass would only increase the $\lya$ background and thus require even lower values for $f_\alpha$. We also did not explore the impact of the star formation efficiency parameter $f_\star$ and the SED. However, for the redshift regime which we explore these parameters are degenerate with $f_\alpha$.

We thus find that heating due to resonance scattering with Lyman series photons may have a significant impact during the Cosmic Dawn and thus should be taken into account when modelling the 21-cm signal. Although we do find that for a fiducial value of X-ray heating ($f_\mathrm{X}=1$), the $\lya$ heating is subdominant, many authors explore a wide range of values for $f_\mathrm{X}$ including low values for which $\lya$ heating will dominate over X-ray heating \citep[e.g][]{2017MNRAS.472.1915C, 2018MNRAS.477.3217G, 2019arXiv190110943M}. We note that none of these papers actually include the effect of $\lya$ heating. 

 Models to explain the absorption feature seen in the EDGES results rely on $\lya$ coupling to produce an observable signal and thus any excess cooling needs to overcome the heating caused by this coupling. As shown in this paper, this pushes for example the millicharged dark matter model into a regime ruled out by other constraints. It remains to be seen if there exist any physically motivated excess cooling processes which can explain the EDGES results.

\section*{Acknowledgements}
The authors would like to thank Hannah Ross, 
Paul Shapiro, Anastasia Fialkov, Sambit Giri, Benedetta Ciardi, Avery Meiksin, Piero Madau,  Tirthankar Roy Choudhury for useful discussions regarding this work. We also like to thank an anonymous referee whose comments have encouraged us to explore the millicharged dark matter model. We have used resources provided by the Swedish National Infrastructure for Computing (SNIC) (proposal number SNIC 2018/3-40) at PDC, Royal Institute of Technology, Stockholm.

\bibliography{bibfile}


\bsp	
\label{lastpage}
\end{document}